# Exploring propane/propylene separation mechanism in ZIF-8 (Zn) by $^2$H NMR spectroscopy.

Khudozhitkov A.E.; Kolokolov D.I.; Stepanov A.G.

**Abstract.**

The molecular mobility of propane and propylene has been studied by $^2$H NMR method. The activation barriers of diffusion determined from the spin relaxation analysis are in a good agreement with the values obtained by Liu et al. ($E_{C3H8}$ = 38 kJ/mol, $E_{C3H6}$ = 13.5 kJ/mol). High activation energy of the rotation inside the cavity for propylene compared to propane (8 kJ/mol vs 3.8 kJ/mol) implies stronger interaction of this adsorbate with the walls of the cavity.

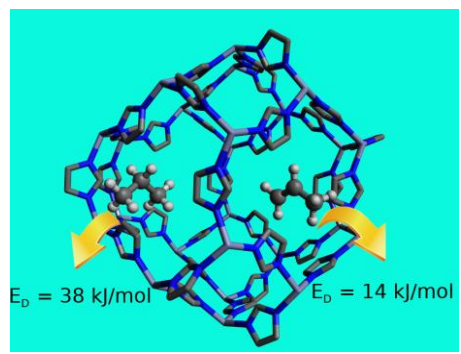

1. Introduction

The separation of propane and propylene is an important task for the chemical industry. Close physic-chemical properties of these molecules make their separation difficult. Currently, the cryogenic distillation process is used despite of its tremendous energy consumption. Therefore, a huge effort has been devoted to the development of the energy efficient methods of separation based on the discrepancy of adsorption and diffusivity properties of propane and propylene.

ZIF-8 is a representative of zeolitic imidazolate frameworks. Its three-dimensional structure consists of large spherical cavities with 11.4 A diameter connected by narrow windows with a size 3.4 A. However, the window aperture is not fixed. Due to the dynamics of the 2-methylimidazolate linkers the effective size of the window increases allowing adsorption of the molecules larger than 3.4 A. Moreover, the flexible nature of the window leads to the high separation selectivity of light hydrocarbons [1-6]. In particular, there are several reports showing the efficiency of ZIF-8 in propane/propylene separation. Several studies have proved that exploitation of ZIF-8 in membrane separation process leads to the outstanding selectivity and permeability values. Pan et al. showed that ZIF-8 membranes deposited on the aluminum support can be successfully used for propane/propylene separation. The separation factor reach up to 50 with the permeability of propylene up to the 200 barrers [7].

Li et al. discovered that although the equilibrium adsorbed amount in ZIF-8 is almost identical for propane and propylene, the diffusivities drastically differ due to the sieving effect. Moreover, the activation barriers 9.7 and 74.1 kJ/mol for diffusion of propylene and propane, respectively, suggest much faster diffusion of propylene [8]. However, the exact values of diffusion activation barriers significantly differ from one research to the other.

Liu et al. studied propene/propylene mixture separation on the ZIF-8 membranes deposited on aluminum support. The separation selectivity was shown to decrease at elevated temperature remaining stable under the various value of external pressure. The activation barriers for diffusion differ from the results obtained by Li et al. The diffusion activation barriers of propylene and propane are 12.7 and 38.8 kJ/mol, respectively [9].

Besides adsorption methods the mobility of propane and propylene in ZIF-8 was investigated by computational methods. For example, Krokidas et al. using molecular dynamics simulations confirmed that the separation of propane/propylene mixture occurs due to the different rates of diffusion instead of thermodynamical preference in the adsorption process. The simulated values of activation barriers of diffusion are 18 and 30 kJ/mol for propylene and propane, correspondingly [10].

Verploegh et al. analyzed the behavior of adsorbed molecules in the transition state near the window. Activation barriers of diffusion derived in this approach are 22.1 and 26.8 kJ/mol for propylene and propane correspondingly [11].

Since, there is a vast spread of reported diffusion parameters of propane and propylene in ZIF-8 we decided to contribute in refining the motional parameters by $^2$H NMR investigation.

2. **Experimental section**

**2.1 Materials.** Nanocrystals of ZIF-8 have been prepared following the recipe of Cravillon et al [12]. First, 0.73 g of $ZnNO_3 \cdot 6H_2O$ (2.45 mmol, 1 equiv) in 50 mL of methanol and 0.81 g of 2-methylimidazole (9.86 mmol, 1 equiv) in 50 mL of methanol were mixed under vigorous stirring and stored for 2 h at room temperature. The precipitate was collected by centrifugation and washed twice with 50 mL of methanol. Further material was dried in nitrogen at 373 K overnight. X-ray diffraction analysis (XRD) proved that it was pure ZIF-8 with a crystal size of 40 nm. SEM image of the nanosized ZIF-8 and the corresponding XRD pattern are given in the Supporting Information.

Predeuterated propane-$d_2$ and propylene-$d_6$ were used in this work.

**2.2 Sample Preparation.** The preparation of the sample for NMR experiments was performed in the following manner. The powder of ZIF-8 (Zn) (~0.06 g) was placed into a special glass cell of 5 mm diameter and 3 cm length. Then the cell was connected to the vacuum line and activated at 523 K for 6 h under vacuum. After cooling the sample back to room temperature, the material was exposed to the vapor of the previously degassed deuterated guest molecules in the calibrated volume (439 cm$^3$) under liquid nitrogen conditions. The quantity of the adsorbed molecules was regulated by the initial vapor pressure created inside the calibrated volume. After adsorption, the neck of the tube was sealed off, while the material sample was maintained in liquid nitrogen to prevent its heating by the flame. Prior to NMR investigations all sealed samples were kept at 423 K for 72 h to allow even redistribution of the guest molecules inside the porous material.

**2.3 NMR Measurements.** $^2$H NMR experiments were performed at the Larmor frequency $\omega_z/2\pi$ = 61.42 MHz on a Bruker Avance-400 spectrometer, using a high probe with 5-mm horizontal solenoid coil. All $^2$H NMR spectra were obtained by Fourier transformation of quadrature-detected phase-cycled quadrupole echoes [13]. Inversion-recovery experiments for the determination of the spin-lattice relaxation time ($T_1$) were performed using the pulse sequence ($180°_x - \tau_v - 90°_{\pm x}$ – acquisition – $t$), where $\tau_v$ was a variable delay between 180° and 90° pulses, $t$ is a repetition delay of the sequence during the accumulation of the NMR signal. The duration of 90° pulse was 1.75 μs. Spin-spin relaxation time ($T_2$) was obtained by a Carr-Purcell-Meiboom-Gill (CPMG) [14] pulse sequence.

The temperature of the samples was controlled with a nitrogen gas flow at low temperatures and air flow at elevated temperatures, stabilized with a variable-temperature unit BVT-3000 with the precision of about 1 K.

**2.4    Modeling.** Modeling $^2$H NMR spectra line shape and spin relaxation rates has been performed with a homemade Fortran program based on the standard formalism applied for description of molecular motions [15,16].

## 3    Results and Discussion

In our previous paper we studied the mobility of xylenes, benzene, toluene and iso-butane in side ZIF-8. Following the same procedure we analyzed the mobility of propane and propylene. $^2$H NMR spectra for both samples exhibit isotropic lineshape even at 113 K, implying the presence of fast isotropic motion (Fig. 1a). However, at high temperature the linewidth decreases and two signals become distinguishable for the propylene sample (Fig. 1b). These signals correspond to 2 kinds of non-equivalent positions in the molecule ($CH_2$ and $CH_3$ groups).

The mobility of adsorbed molecules was studied from the analysis of the spin relaxation temperature dependence. Although, the presence of two different species of deuterium in propylene molecules could make the analysis more difficult, Fig.2 shows that the spin relaxation times of unresolved signal at low temperature smoothly follow the trend of the methyl group signal that has higher intensity at high temperature. Therefore, we can conclude that this signal dominates at low temperature and governs the relaxation. That is why in the modeling of the experimental data for propylene we treated this signal as the signal of methyl group.

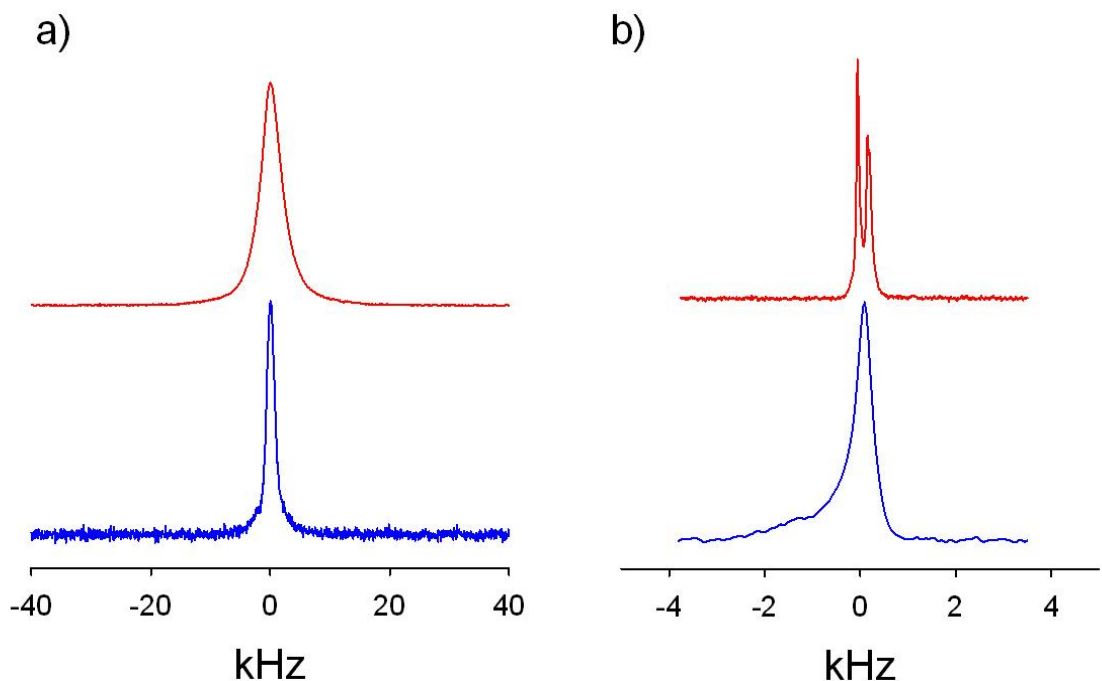

Figure 1. $^2$H NMR spectra of propane (blue lines) and propylene (red lines) at 113 K (a) and at high temperature limit (b). [513 K for propane, 363 K for propylene]

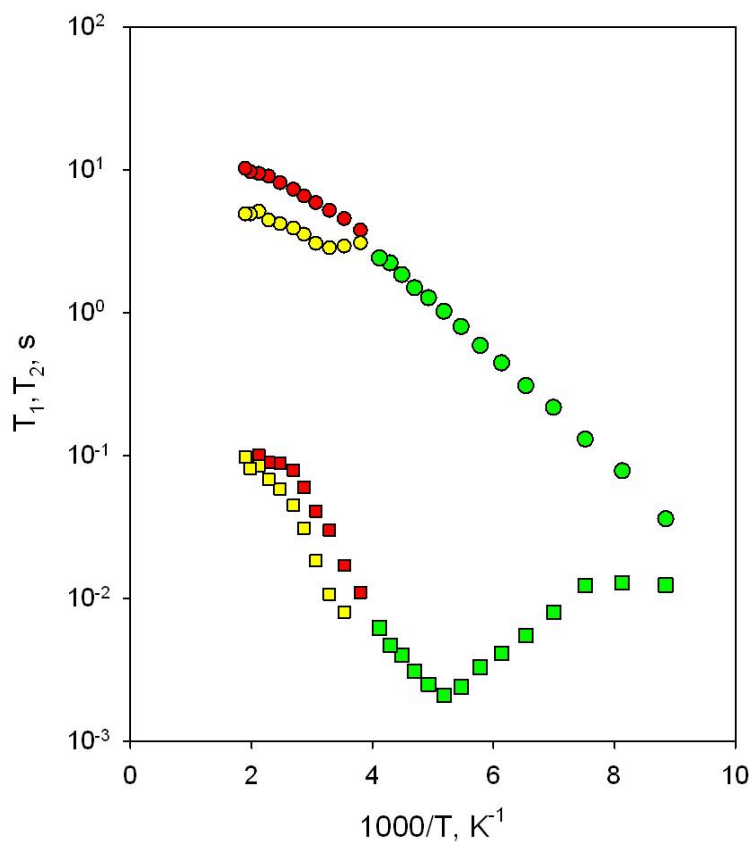

Figure 2. Spin relaxation times of propylene as a function of the temperature. Green symbols represent the relaxation of the unresolved signal, red symbols represent the relaxation of the more intense signal, and yellow symbols represent the relaxation of the weaker signal.

The spin-spin relaxation time ($T_2$) of propylene has a minimum with the following steep raise at elevated temperature. From the adsorption studies it is known that propylene rapidly diffuse through the ZIF-8, so this high energetic motion probably corresponds to the diffusion. Confidence in that conclusion becomes even stronger after comparing the activation barrier of the motion that governs the diffusion with the results based on the sorption kinetics through membrane. Activation energy 13.5 kJ/mol is close to the value 12.7 kJ/mol measured by Liu et al. Moreover, if we assume the length of the diffusion jump equal to the size of the pore than the rate of diffusion will be reproduced as well ($D_0 \sim 3 \times 10^{-6}$ cm$^2$/s from our results and $2 \times 10^{-6}$ cm$^2$/s from Liu et al.).

However, the activation of diffusion would have to lead to the same values of $T_1$ and $T_2$ relaxation in the high temperature limit. It is not the case as can be seen from the experimental data. Therefore, following our previous work we assume the presence of two populations of the guest molecules: one is constituted by the molecules moving freely in the cavity (A), second represents molecules localized near the windows from the cavity (B). Mobile species govern the $T_1$ relaxation, whereas the molecules localized near the windows affect mostly $T_2$ relaxation and lead to the discrepancy between $T_1$ and $T_2$ relaxation times in the high temperature region.

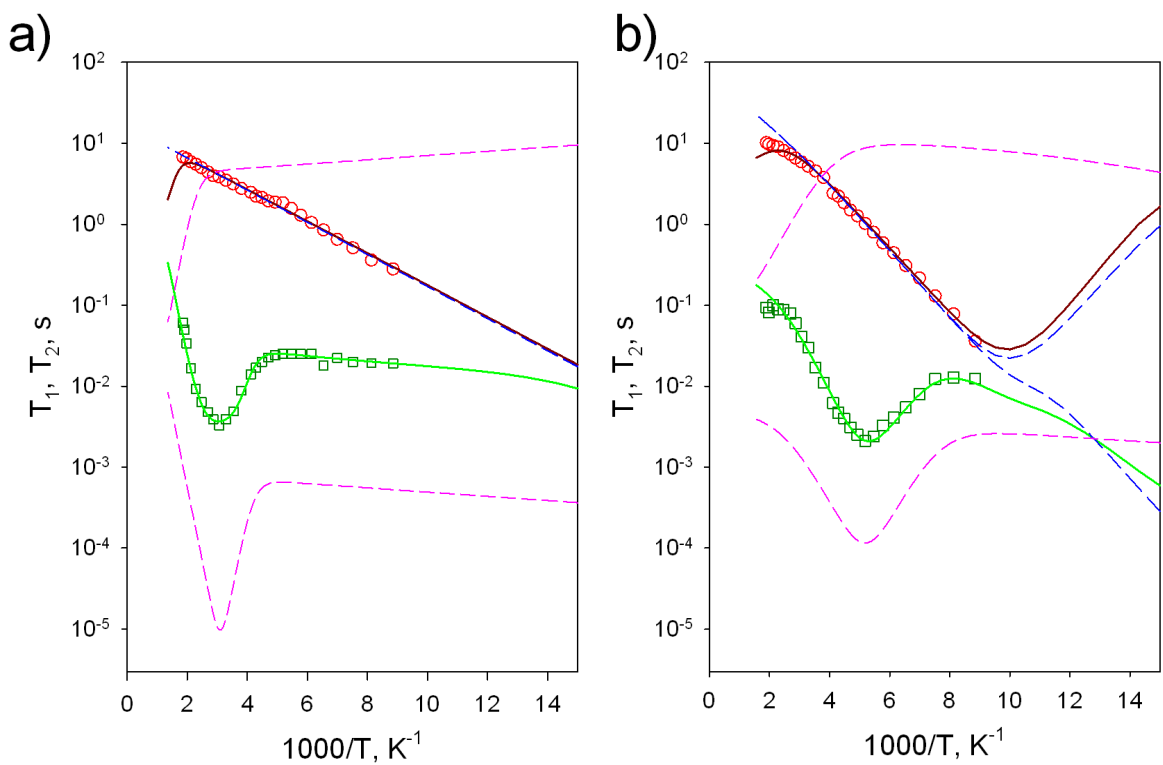

Figure 3. Temperature dependence of the $T_1$ (red circles) and $T_2$ (green squares) spin relaxation times for propane (a) and propylene (b). Individual relaxation times are presented with dashed lines for A (blue) and B (pink) species. Final relaxation curves after exchange between these populations are presented with solid lines.

Fig.3 shows experimental values of spin relaxation times as well as the fitting for both samples. Minimum on the $T_2$ curve for propane occurs at higher temperature (by 130 K) than for propylene, whereas the slope of the relaxation curve after minimum is steeper. Such behavior is expectable since this motion corresponding to the diffusion and propane is known to have slower rate and higher activation barrier of diffusion than propylene. Activation barriers of diffusion from our experiment turned out to be 38 kJ/mol and 13.5 kJ/mol for propane and propylene respectively. These values are close to the results of Liu obtained from the membrane breakthrough experiment 38.8 and 12.7 kJ/mol for propane and propylene respectively.

At the same time there are unexpected results as well. Basing on the slope of $T_1$ relaxation curve we can conclude that isotropic rotation of propylene inside the cavity is more hindered than the rotation of propane. Activation barrier for that motion in case of propane is 8 kJ/mol that is almost twice as high as the barrier for propane 3.8 kJ/mol. It seems that slightly larger size of propane molecule decelerates the rotation weaker than the interaction of double bond of propylene with the framework. So it is harder for propylene to rotate inside the cavity whereas the transport between pores occurs faster than for the propane.

All motional parameters used in the modeling are presented in Table 1.

Table 1. Motional parameters used in fitting the spin relaxation curves.

|  | propane | propylene |
|---|---|---|
| $E_{iso}$, kJ/mol | 3.8 | 8 |
| $k_{iso}$, Hz | $1.2 \times 10^{12}$ | $1.5 \times 10^{12}$ |
| $E_D$, kJ/mol | 38 | 13.5 |
| $k_D$, Hz | $1.5 \times 10^{10}$ | $4.5 \times 10^{8}$ |
| $E_{lib}$, kJ/mol | 0.5 | 0.5 |
| $k_{lib}$, Hz | $10^6$ | $10^6$ |
| $E_{CH3}$, kJ/mol | – | 1.5 |
| $k_{CH3}$, Hz | – | $5 \times 10^{12}$ |
| $E_{ex}$, kJ/mol | 1.5 | 1.5 |
| $k_{ex}$, Hz | $3 \times 10^3$ | $10^5$ |
| $E_{eq}$, kJ/mol | 0.5 | 2.5 |
| $K_{eq}$ | 0.023 | 0.014 |
| $\theta$ | 110 | 120 |
| $\xi$ | 54.7 | 20 |

## 4 Conclusion

The molecular mobility of propane and propylene has been studied by $^2$H NMR method. The activation barriers of diffusion determined from the spin relaxation analysis are in a good agreement with the values obtained by Liu et al ($E_{C3H8}$ = 38 kJ/mol, $E_{C3H6}$ = 13.5 kJ/mol). High activation energy of the rotation inside the cavity for propylene compared to propane (8 kJ/mol vs 3.8 kJ/mol) implies stronger interaction of this adsorbate with the walls of the cavity.

**Acknowledgements.**

This work was supported by the Russian Foundation for Basic Research (grant no. 18-33-00048).

**References.**

(1) Diestel, L.; Bux, H.; Wachsmuth, D.; Caro, J. *Microporous Mesoporous Mater.* **2012**, *164*, 288.

(2) Pérez-Pellitero, J.; Amrouche, H.; Siperstein, F. R.; Pirngruber, G.; Nieto-Draghi, C.; Chaplais, G.; Simon-Masseron, A.; Bazer-Bachi, D.; Peralta, D.; Bats, N. *Chem. Eur. J.* **2010**, *16*, 1560−1571.

(3) Peralta, D.; Chaplais, G.; Simon-Masseron, A.; Barthelet, K.; Chizallet, C.; Quoineaud, A. A.; Pirngruber, G. D. *J. Am. Chem. Soc.* **2012**, *134*, 8115.

(4) Peralta, D.; Chaplais, G.; Paillaud, J. L.; Simon-Masseron, A.; Barthelet, K.; Pirngruberb, G. D. *Microporous Mesoporous Mater.* **2013**, *173*, 1.

(5) Zhang, K.; Lively, R. P.; Zhang, C.; Chance, R. R.; Koros, W. J.; Sholl, D. S.; Nair, S. *J. Phys. Chem. Lett.* **2013**, *4*, 3618.

(6) Chang, N.; Gu, Z. Y.; Yan, X. P. *J. Am. Chem. Soc.* **2010**, *132*, 13645.

(7) Pan, Y. C.; Li, T.; Lestari, G.; Lai, Z. P. *J. Membr. Sci.* **2012**, *390*, 93.

(8) Li, K. H.; Olson, D. H.; Seidel, J.; Emge, T. J.; Gong, H. W.; Zeng, H. P.; Li, J. *J. Am. Chem. Soc.* **2009**, *131*, 10368.

(9) Liu, D. F.; Ma, X. L.; Xi, H. X.; Lin, Y. S. *J. Membr. Sci.* **2014**, *451*, 85.

(10) Krokidas, P.; Castier, M.; Moncho, S.; Brothers, E.; Economou, I. G. *J. Phys. Chem. C* **2015**, *119*, 27028.

(11) Verploegh, R. J.; Nair, S.; Sholl, D. S. *J. Am. Chem. Soc.* **2015**, *137*, 15760.

(12) Cravillon, J.; Munzer, S.; Lohmeier, S. J.; Feldhoff, A.; Huber, K.; Wiebcke, M. *Chem. Mat.* **2009**, *21*, 1410.

(13) Powles, J. G.; Strange, J. H. *Proc. Phys. Soc. London* **1963**, *82*, 6.

(14) Farrar, T. C.; Becker, E. D. *Pulse and Fourier Transform NMR. Introduction to Theory and Methods*; Academic Press: New York and London, 1971.

(15) Wittebort, R. J.; Olejniczak, E. T.; Griffin, R. G. *J. Chem. Phys.* **1987**, *86*, 5411.

(16) Macho, V.; Brombacher, L.; Spiess, H. W. *Appl. Magn. Reson.* **2001**, *20*, 405.